\documentclass[12pt,preprint]{aastex}

\newcommand{\sax}{{\it BeppoSAX}}

\newcommand{\hete}{{\it HETE-2}}

\shorttitle{Long duration component from short GRBs}
\shortauthors{Montanari et al.}

\begin{document}

\title{Evidence for a long duration component in the prompt emission of
short Gamma--Ray Bursts detected with \sax}

\author{Enrico Montanari\altaffilmark{1,2},
         Filippo Frontera\altaffilmark{1,3}, 
          Cristiano Guidorzi\altaffilmark{1,4},
          and Massimo Rapisarda\altaffilmark{5}}

\altaffiltext{1}{Dipartimento di Fisica, Universit\`a di Ferrara, Via Paradiso 12, 
44100 Ferrara, Italy; montanari@fe.infn.it}
\altaffiltext{2}{ISA ``Venturi'', Modena, Italy}
\altaffiltext{3}{IASF, CNR, Via Gobetti 101, 40129 Bologna, Italy}
\altaffiltext{4}{Now at Astrophysics Research Institute, Liverpool John Moores University, UK}
\altaffiltext{5}{ENEA Divisione Fusione, Centro Ricerche di Frascati,
  CP 65, 00044 Frascati, Rome, Italy}

\begin{abstract}
A statistical study on the light curves of all the short  
Gamma--Ray Bursts detected with the Gamma Ray Burst Monitor (GRBM)
aboard \sax\ is reported.
Evidence for a very weak and long duration component associated with
these events in the two 1 s counters of the  GRBM (40-700 keV and $>$100 keV)
is found. It starts a few tens of seconds before the burst and 
continues for about 30 s after the burst.
The overall hardness of this component is comparable with that of the 
event itself. The detection of a signal before the onset time and 
the similar hardness are consistent with an interpretation of the 
long duration component in
terms of prompt emission associated with short GRBs. 

\end{abstract}

\keywords{gamma rays: bursts --- methods: statistical}

\section{Introduction}

The duration distribution of Gamma Ray Bursts (GRBs), 
discovered by BATSE (Kouveliotou et al. 
1993\nocite{kouveliotou93}) and confirmed
by \sax\ (Guidorzi 2002, Guidorzi et
al. 2004\nocite{guitesi02,guidorzi04}), 
shows a bimodal
pattern with two peaks. 
This distribution divides GRBs into 
long 
($>2$~s) and short 
($<2$~s) GRBs. A well established feature of short GRBs is
their spectral hardness, which appears significantly higher than that of
long GRBs (Kouveliotou et al. 1993, Belli 
1995\nocite{kouveliotou93,belli95}). A convincing physical 
interpretation for this bimodal distribution is still lacking. An
hypothesis is that short GRBs could have origin from the merging of 
compact stars, while long GRBs from the collapse of massive stars 
(see, e.g., review by Zhang and M\'esz\'aros 2004\nocite{zhang04}).
Some unified models that might account for the
bimodal distribution have recently been proposed (e.g. Yamazaki et al. 
2004, Toma et al. 2005\nocite{yamazaki04,toma05}).
As it occurred in the case of long GRBs, the mystery of short GRBs could
possibly be unveiled by the discovery of their counterparts at lower 
frequencies (X--ray, optical, radio). However, so far all these searches 
have been unsuccessful. 
In X-rays, no short GRB has been detected with the \sax\ Wide Field Cameras
(Smith et al. 1999\nocite{smith99}) and the 2 short GRBs detected 
with \hete\ (GRB020113,
Barraud et al. 2003\nocite{barraud03}; GRB020531, Ricker et al. 
2002\nocite{ricker02}), have provided unsuccessful searches for 
optical afterglow, in particular for GRB020531, which was followed up 
from ~90 min after the event (Boer et. al 2002\nocite{boer02}) 
until 10 days later
(for a review of the observations see Klotz et al. 2003\nocite{klotz03}). 
From these observations it appears that, if present, the afterglow
from GRB020531 should be either much weaker than the weakest afterglow observed from
the long GRBs, or it should be of very short duration or it would show 
a very fast fading. The search for afterglow emission from short GRBs 
is one of the hottest topics of the current astrophysical research, 
of key importance to establish the origin of these still mysterious events.
From these considerations it is apparent that even a statistical study 
of this class of events aimed at possibly discovering the presence of a 
long duration component, which could be reminiscent of a hard X--/$\gamma$--ray afterglow,
would be of great value. In fact Lazzati et al. (2001)\nocite{lazzati01}
and Connaughton (2002) were the first to perform such analysis using BATSE data. 
Connaughton (2002) added together
100 background-subtracted $>$20 keV light curves of short GRBs, aligning them at
the peaks of the events, and finding a weak signal (3.5$\sigma$ significance)
in the time interval from 20 to 40 s after the main event and a marginal
signal later (until 300 s). Lazzati et al. (2001) 
for each of the 4 BATSE energy channels, added together the 
light curves of 76 short GRBs with high signal-to-noise
ratio, and found evidence for a long duration signal (at $\sim 4\sigma$ significance) 
above the mean background level in the 25-120 keV energy band (sum of
the channels 1 and 2). 
In both cases the signal starts soon after the bursts; in the case of Connaughton
(2002) it seems to be fading, while in the case of Lazzati et al. (2001), 
it peaks about 30~s after the burst time and it is visible for about 100~s.
Lazzati et al. (2001) did  not find any signal above 120 keV. 
Lazzati et al. (2001) also found that the energy spectrum of the 25--120 keV excess 
appears to be softer than that of the short GRBs, and suggested 
that it could be due to afterglow emission from these events.
Here we report on a statistical analysis similar to that performed by Lazzati
et al. (2001), using the light curves of the short GRBs detected by the \sax\ 
Gamma Ray Burst Monitor (GRBM). 
All these bursts will appear in the GRBM catalogue of 
GRBs now in preparation, with preliminary information already reported
(Guidorzi 2002, Guidorzi et al. 2004\nocite{guitesi02,guidorzi04}). 
Preliminary results of this analysis, obtained with a smaller number
of events, have already been reported 
(Montanari et al. 2004\nocite{montanari04}).

\section{Data analysis and results}

\begin{figure}[!t]
\resizebox{6in}{!}{\includegraphics[angle=-90]{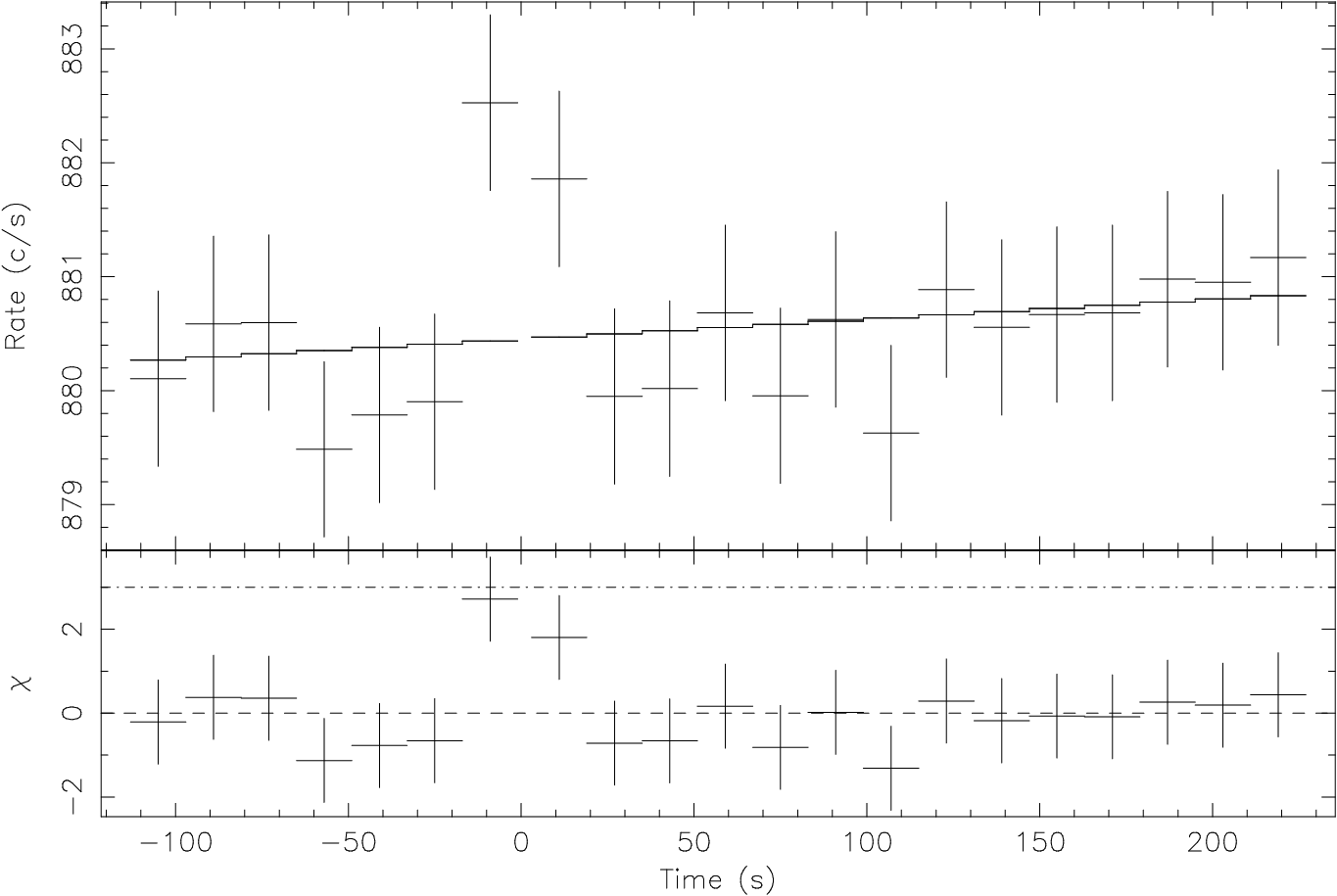}}
\resizebox{6in}{!}{\includegraphics[angle=-90]{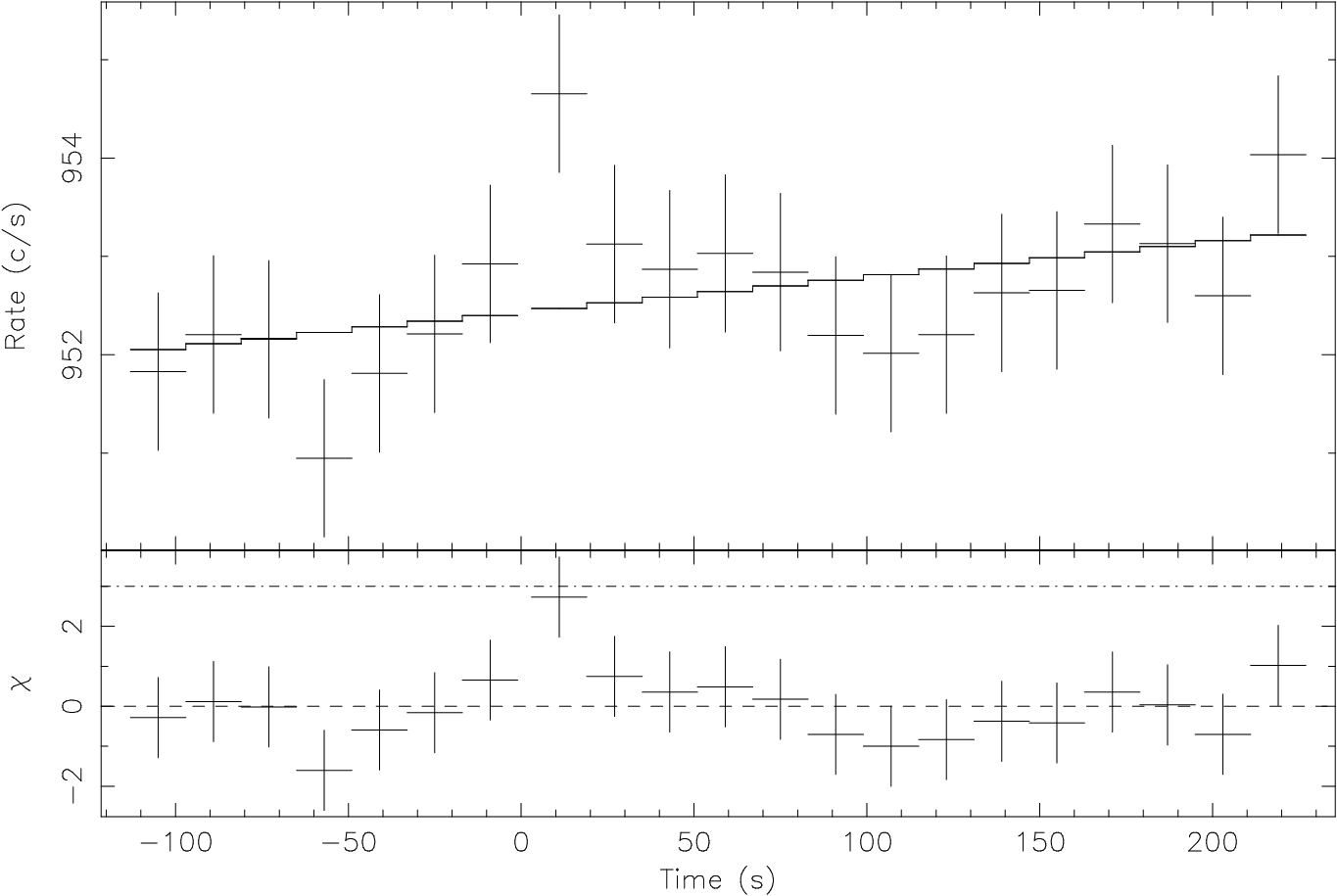}}
\caption{Average light curve around the trigger time 
(0 of the Time axis) of 93 short GRBs. The best fit points
obtained by fitting the data with a $1^{st}$ degree polynomial are shown.
{\em Top:} 40--700 keV; {\em bottom:} $>$100 keV.
}

\label{f:int}
\end{figure}

\begin{figure}[!t]
\resizebox{6in}{!}{\includegraphics[angle=-90]{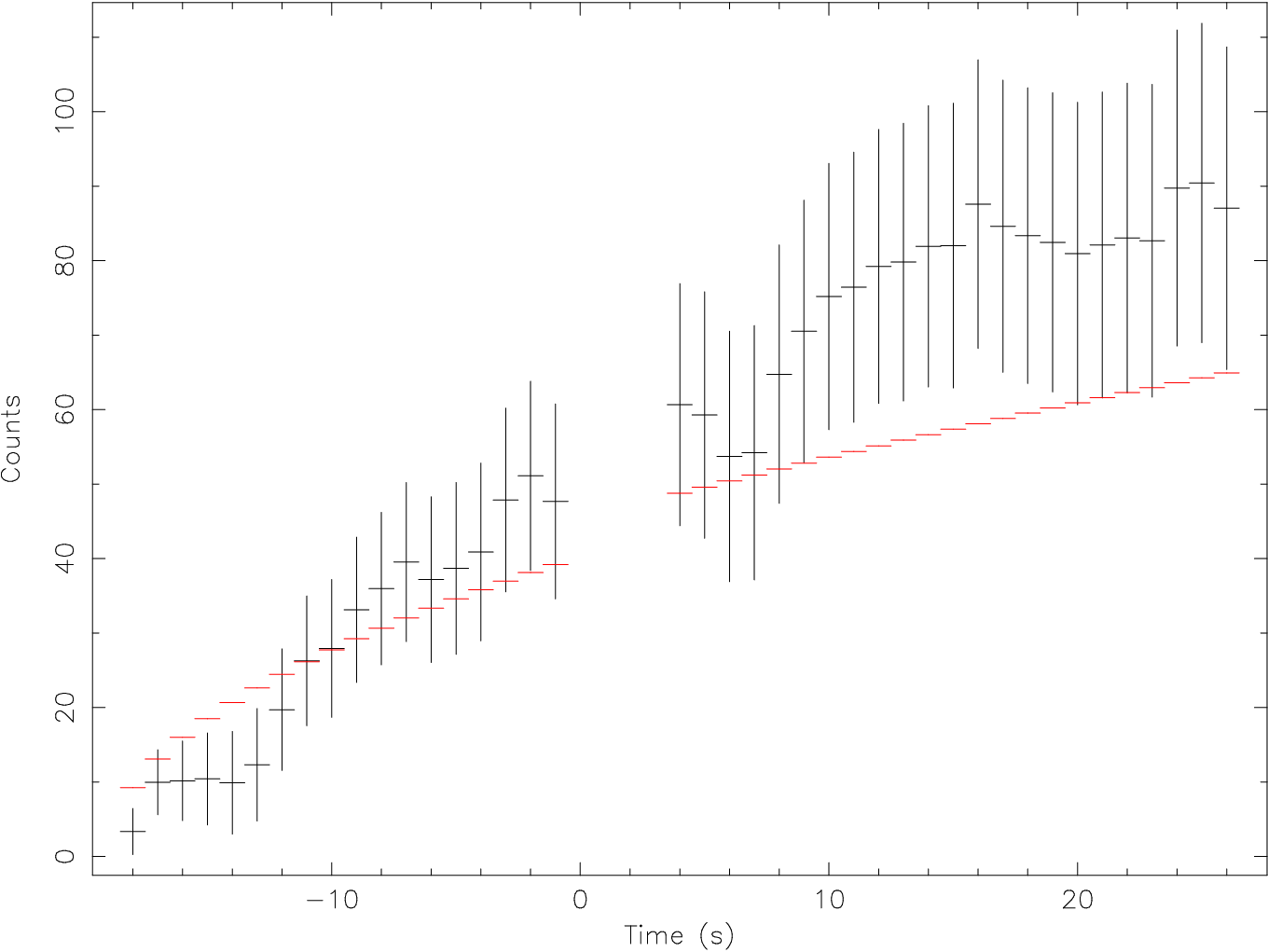}}
\resizebox{6in}{!}{\includegraphics[angle=-90]{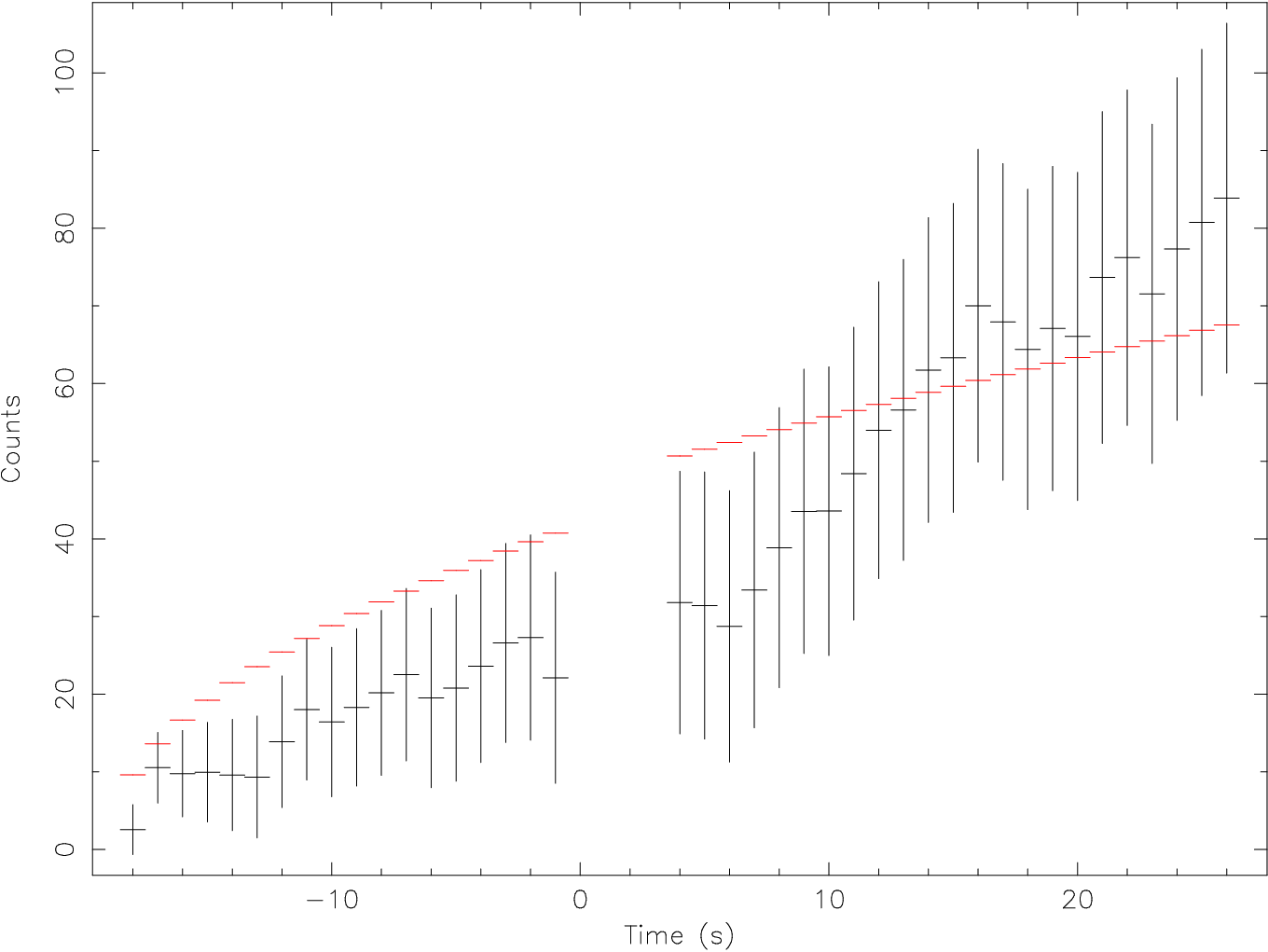}}
\caption{Cumulative counts per burst of the average
background-subtracted light curves vs. time from $-19\,{\rm s}$ assumed
as origin of the Time axis. Red marks gives the $3\,\sigma$ level of
the cumulative excess.
{\em Top:} 40--700 keV. {\em Bottom:} $>$100 keV.
}
\label{f:cum}
\end{figure}

The main features of the \sax\ GRBM have been given in many papers 
(Frontera et al. 1997, Feroci et al. 1997, Costa et al. 
1998\nocite{frontera97,feroci97,costa98}). 
The instrument is made of four 
CsI(Na) scintillator detection units with 
an energy 
passband from 40 to 700 keV for each unit.
Data continuously available from each of the 4 GRBM units include 
two 1~s ratemeters in two energy
channels (40--700~keV and $>$100~keV), and 128~s count spectra
(40--700~keV, 225 channels). 
The number of short GRBs detected by the \sax\ GRBM is 168.
For each of them we have used the 1 s light curves obtained
from the GRBM detection unit which showed the maximum count rate.
Thanks to the almost equatorial \sax\ orbit, the GRBM background level
is much more stable than that of BATSE, with a maximum orbital
modulation of about 15\%, and an average intensity level of about 
800 counts/s. 
First of all we aligned the GRBs light curves with respect to the trigger time
in such a way that the detectable prompt emission for each GRB was located in the interval
from $-1$~s to $3$~s. 
After that we selected those events for which no interruption
in the data transmission was present in either energy channel (40--700 keV and
$>$100 keV) from $-113$~s to $+227$~s  (roughly the same time interval chosen
by Lazzati et al. 2001) with respect to the GRB trigger time.
We rebinned these data in 21 time bins of 16~s duration: 7 bins from 
$-113$~s to $-1$~s and 14 from $+3$~s to $+227$~s.
We fit each of the resulting light curves with a $1^{\rm st}$
degree polynomial and rejected those GRBs for which the corresponding
light curve was not fit by the above function at a significance level less
than 0.01 in at least one of the two energy channels. With this procedure, 
the GRB sample was reduced to 117 events.
In order to exclude variability at the $1\,{\rm s}$ timescale that can
be possibly smoothed away at the $16\,{\rm s}$ timescale,
we took the previously selected events
and fit each of the original $1\,s$ light curves in the interval
$[-113,-1]\cup[3,227]\,s$ with a $1^{\rm st}$
degree polynomial, rejecting those GRBs for which the corresponding
light curve was not fit by the above function at a significance level less
than 0.01 in at least one of the two energy channels. With this procedure, 
the GRB sample was reduced to 93 events.
The resulting $1\,{\rm s}$ average light curves in the two energy 
bands show a significant excess lasting some tens of seconds around
 the trigger time. For the
40--700 keV band, a $4\,\sigma$ maximum excess is found in the interval
$[-19,-1]\cup[3,16]\,{\rm s}$, while for the $>$100 keV band a $3.3\,\sigma$ 
significant excess in the $[-19,-1]\cup[3,26]\,s$ interval is detected.
Most importantly, in the 40--700 keV energy band,
an almost $3.7\,\sigma$ significant excess is found before
the onset of the main short GRB emission in the interval
$[-19,-2]\,s$. 
In the $>$100 keV energy band, an excess of
$3.6\,\sigma$ significance level is found only after the main short GRB 
emission ($[6,26\,s$). Summarizing, a signal is found both before and
after the short GRB emission, with evidence of a spectral evolution 
from soft to hard.
The resulting time behaviour of the  average light curves rebinned at $16\,{\rm s}$
is shown in Fig.~\ref{f:int}, along with the fit with a
$1^{\rm st}$ degree polynomial and the distribution of the residuals from 
the best fit function. As can be seen, the mean background level is
well described by this function apart from the excess in the time 
intervals contiguous to that of the short GRB emission. 
Excluding the interval $[-17, 19]\,$s from the fit, 
the excess becomes even more robust, as can also be seen in
Fig.~\ref{f:hnt}. It is found that the interval for which the
maximum Signal to Noise Ratio (SNR) is reached in both energy bands is 
${\cal C}=[-19,-1]\cup[3,26]\,$s ($4.0\,\sigma$ and $3.7\,\sigma$ for
the 40--700 keV and $>$100 keV band, respectively). The maximum SNR
before the onset of the prompt emission ($4\,\sigma$) is detected in the
interval ${\cal A}=[-19,-2]\,$s for photons in the 40--700 keV band. 
After the prompt emission, the time interval with the
highest SNR in both energy bands is ${\cal B}=[6,16]\,$s 
with $3.3\,\sigma$ in the 40--700 keV
band and and $3.7\,\sigma$ for photon energies $>$100 keV. 
For a summary of the results see Table~\ref{tab1}.
Overall, the excess has the appearance of an almost symmetrical
bump (see Fig.~\ref{f:hnt}), and can be described by a Gaussian function 
centred at the time origin, with a F--test significance of $1.8\cdot 10^{-3}$ for
the 40-700 keV energy band and $6.2\cdot 10^{-3}$ for the higher passband. This result was
obtained by fitting the average light curves rebinned at $8\, $s with
a Gaussian
centered at the origin plus a $1^{st}$ degree polynomial.
The shape and significance of the excess is better shown in Fig.~\ref{f:cum},
in which we plot the cumulative counts of the  background subtracted 
40--700 and $>100 keV$ $1\,{\rm s}$ light curves versus time, starting from $t_0 = -19\,{\rm s}$ 
assumed as origin. As can be seen, in the 40--700 keV band the cumulative counts 
steadily and significantly increase with time, while at energies $>$100 keV, the increase
starts to be significant only after the short GRB.

\section{An additional reliability test of the found excess}

The reliability of the obtained results was also tested by repeating the above 
analysis with light curves which did not include short GRBs (we termed them
``background light curves'').
We obtained from the GRBM data archive a set of 1409 pairs (one for
each energy channel) of background light curves
spanning a time interval from $-113$~s to $-1$~s and from $+3$~s 
to $+227$~s with respect to a randomly chosen time origin.
From these light curves, we randomly sorted 10 sets of 93 pairs of
light curves, each set being statistically 
equivalent to that of the light curves derived for the final GRB sample
discussed above. Also in this case we have obtained for every set two
average light curves (one for each energy band). 
All average background light curves are well fit by a $1^{st}$ degree
polynomial. 
Besides, the fluctuations of the curves with respect to the fit 
(given in standard deviation units) are all consistent with a Gaussian,
with zero mean and unit standard deviation.
On timescales of few tens of seconds, no significant excess is
found in any part of the light curves.
For a $40\,$s binning time, the maximum excess found is at a
$2.4\,\sigma$ level and not simulataneously in both energy channels.

\section{Spectral analysis}

In order to understand the nature of the found excess, we compared its spectral 
properties with the mean properties of the 93 short GRBs in our final sample.
We investigated spectral hardness, energy flux and fluence of 
both the long duration component and main event.
To this aim we used the response functions of the  four GRBM units, which were 
derived with Monte Carlo techniques and were tested up to 1
MeV (e.g., Calura et al. 2000\nocite{calura00}).
The mean response function was 
obtained weighting that
of each GRBM unit by the fraction of the GRB events whose light curve
was extracted from that unit.  
The mean count $C$ per event of the 93 short GRBs in our sample
was evaluated in two ways, either averaging the background subtracted
counts derived per each event, or using the average light curve of
the 93 GRBs. In the first case we obtained $C(40-700)=471\pm5$ 
counts/grb and $C(>100)=383\pm5$ counts/grb, for the
40--700 keV and the $>$100 keV energy interval, respectively,
while, in the second case, $C(40-700)=457\pm7$ counts/grb and 
$C(>100)=374\pm7$~counts/grb.
We see that in the second case there is a slight trend to understimate
the total counts, in particular in the low energy band. This causes a slight
increase of the hardness ratio $HR = C(>100)/C(40-700)$. However, since 
the excess around the trigger time can be only evaluated in the second way, 
for consistency reasons, we used the last method also for estimating the mean 
GRB total counts (see Table~\ref{tab1}). The benefit is that,
using the average light curve of the 93 GRBs, the excess is not 
overestimated and its hardness ratio is not underestimated. $HR$ for
both main event and long duration component is reported in Table~\ref{tab1}.
The mean count of the excess normalized to a single event in the
two  energy channels
was evaluated using the best fit light curves shown in 
Fig.~\ref{f:hnt}, and is reported in Table~\ref{tab1} for the time 
intervals ${\cal A}=[-19,-2]\,{\rm s}$,
${\cal B}=[6,16]\,s$ and ${\cal C}=[-19,-1]\cup[3,26]$.
The first interval has a duration of $17\,{\rm s}$ before the GRB onset, 
the second a duration of $10\,{\rm s}$ after the GRB, while the duration of
the last is 41 s.
We fit the total counts of both short events and excess with a 
power--law model ($N(E) \propto E^{-\Gamma}$). 
This is an approximate model also for short GRBs in the 40-700
keV band (e.g., Ghirlanda et al. 2004\nocite{Ghirlanda04}), even though in some
cases (e.g., GRB020531, Lamb et al. 2004\nocite{Lamb04}) this model 
suffices to give
a good description. In our case, with two energy  channels this is the best 
description of the spectrum we can do. 
However, the uncertainty in the mean response function derived, does
not allow a better accuracy.
The best fit value of $\Gamma$ for the average GRB emission 
is reported in Table~\ref{tab1}, where we also report the hardness ratio, 
and mean flux and fluence per burst. As can be
seen, the derived value of $\Gamma$ is consistent with that found in other 
short GRBs (see, e.g., Paciesas et al. 2001\nocite{Paciesas01}, Lamb et al. 
2004\nocite{Lamb04}).  
In the same Table~\ref{tab1} we also report the 40--700 keV fluence
$S(40-700)$ and the mean flux $F(40-700)$ of the long duration excess
found, normalized to a single GRB.  As can 
be seen, within the uncertainties no significant change of the flux and fluence from
the interval before the burst to that after the burst can be
inferred. However, from the derived values of either the $HR$ or the photon index 
$\Gamma$, it appears that the excess before the burst is softer than that
after the burst at a significance level of $1.2\cdot10^{-3}$ ($>3\,\sigma$).
The best estimate of the parameters reported in Table~\ref{tab1}  
is obtained in the interval ${\cal C}=[-19,-1]\cup[3,26]$. In
this interval, it appears that the mean photon index $\Gamma$ of 
the long duration component ($0.2^{+0.6}_{-1.0}$) 
results to be consistent with that of the main event
($1.30\pm 0.06$) at a significance level $3.3\cdot 10^{-2}$ ($2.1\,\sigma$).
In this time interval also the mean energy flux and fluence are well 
determined. 
The 40--700 keV 
energy flux of the long duration component is only $\sim 1\%$ of the
GRB mean flux, while its fluence is almost comparable 
($\sim 20\%$) to that of the main event.

\section{Discussion}

\begin{figure}[!t]
\resizebox{6in}{!}{\includegraphics[angle=-90]{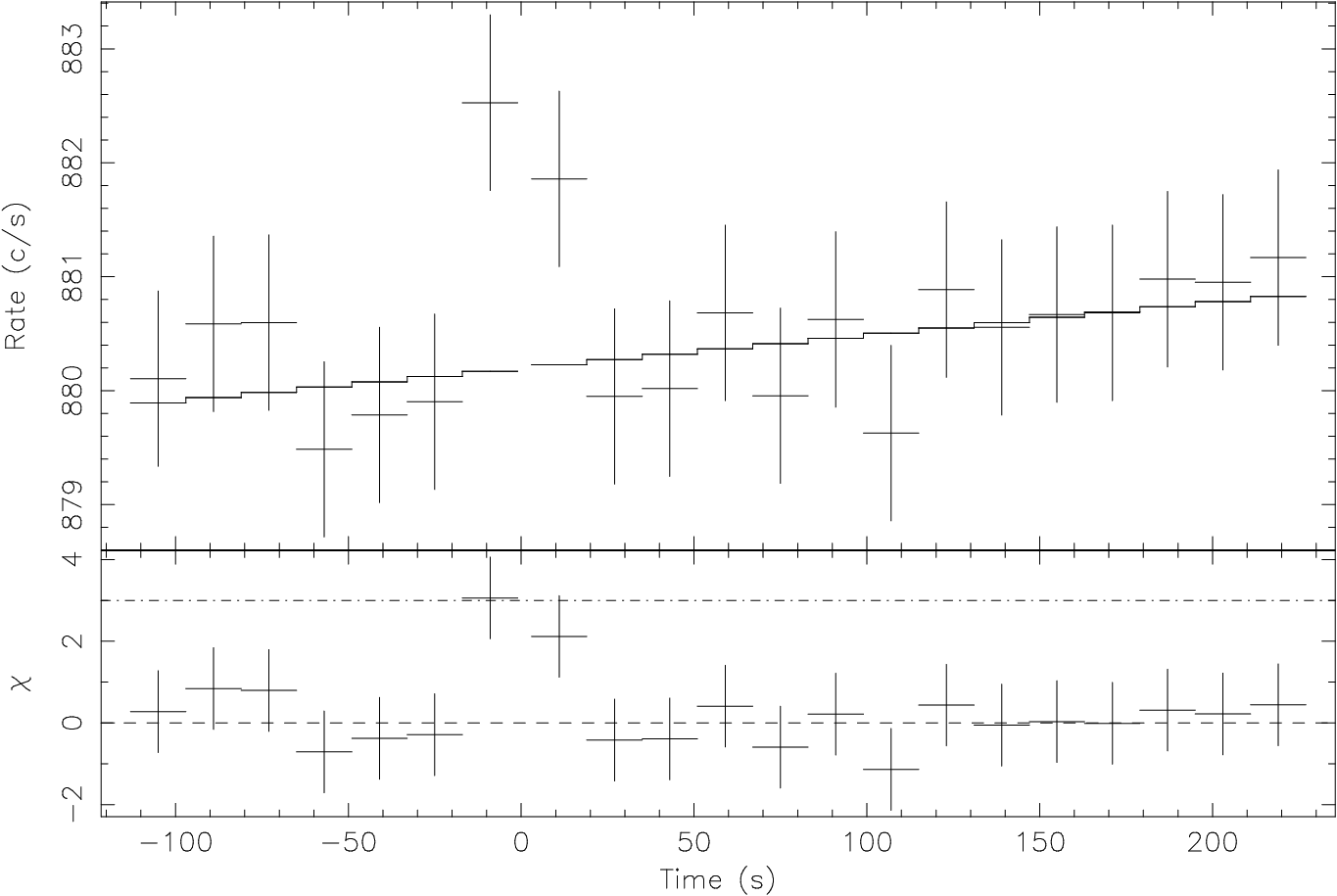}}
\resizebox{6in}{!}{\includegraphics[angle=-90]{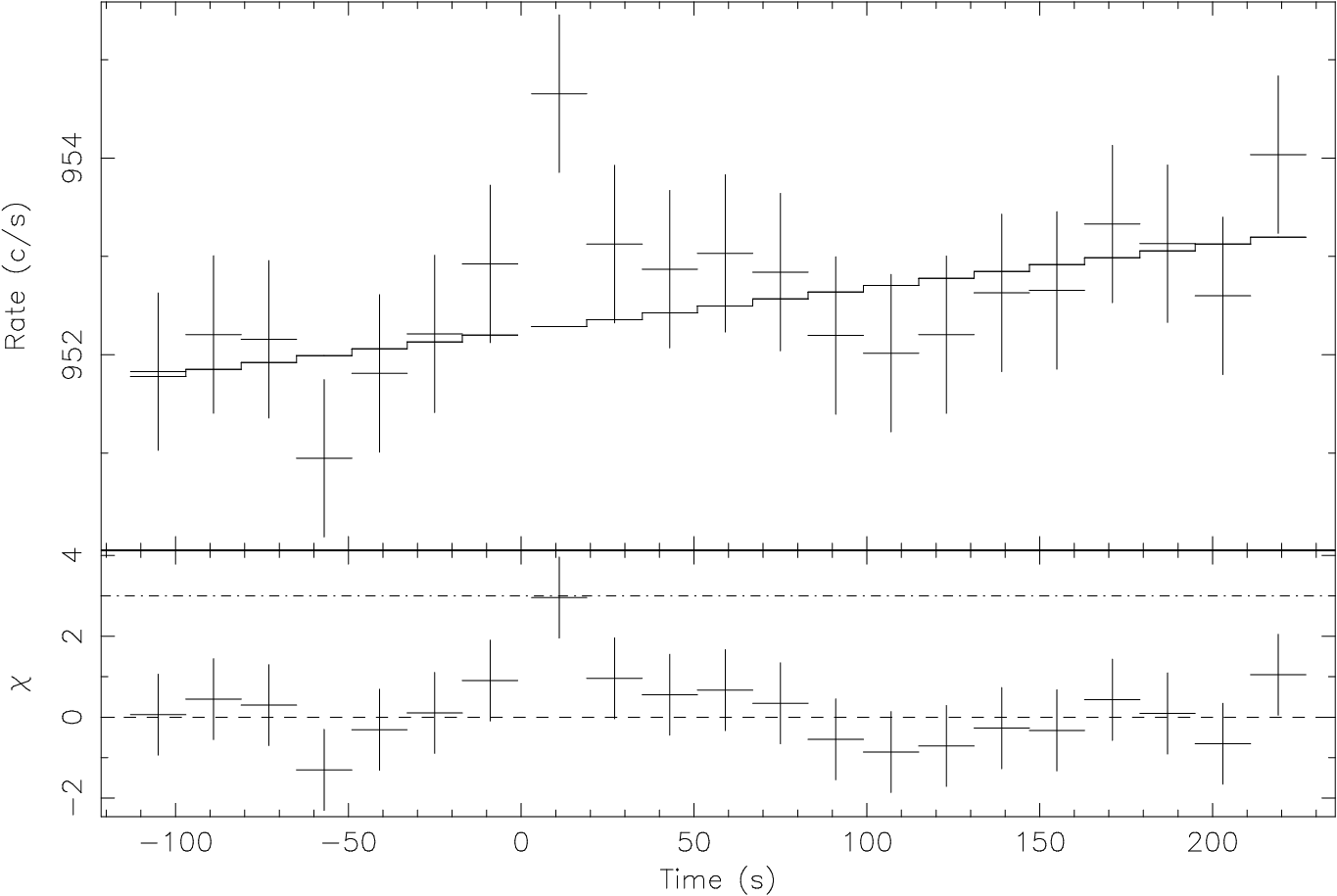}}
\caption{Average light curve around the trigger time 
(0 of the Time axis) of 93 short GRBs. The best fit points
obtained by fitting, with a $1^{st}$ degree polynomial, the light curve 
data out of the $(-17,19)\,{\rm s}$ time interval are shown. 
{\em Top:} 40--700 keV; {\em bottom:} $>$100 keV.
}

\label{f:hnt}
\end{figure}

The \sax\ GRBM  data in the 40--700 keV and $>$100 keV energy channels 
confirm the evidence, initially reported by 
Lazzati et al. (2001)\nocite{lazzati01} 
(in the 25--110 keV band) and  Connaughton (2002) (at energies $>20$~keV) 
using the BATSE data, for a long duration component associated with 
short GRBs, starting soon after the short event.
The 40--700 keV flux in our interval ${\cal B}$ (after the short
event) is in good agreement with that reported by Lazzati et al. 
(2001)\nocite{lazzati01}
and also the peak flux found by these authors at 50 keV 
($\sim 10^{-11}$~erg~cm$^{-2}$~s$^{-1}$~keV$^{-1}$) is consistent with
that derived from our spectral data.
However our results differ from those obtained with the BATSE data.
First, unlike Lazzati et al. (2001)\nocite{lazzati01}, 
we find a significant 40--700 keV count 
excess above the background level also before the trigger time 
(see Fig.\ref{f:hnt}). As can be seen from the figure, the overall excess 
has the appearance of a symmetrical bump, well described 
by a Gaussian function centred at the time origin (see the end of
Sec. 2).
Second, after the short GRB we find a clear signal above 100 keV, 
which is not observed in the BATSE data analyzed by Lazzati et al. (2001)\nocite{lazzati01}. 
As a consequence of this high energy signal, we find that the mean spectral hardness 
of the excess (see Table~\ref{tab1}) is consistent with that of the short events, 
unlike that reported by Lazzati et al. (2001)\nocite{lazzati01} which
is 
softer.
These seemingly different results are likely due to the different
sample of short GRBs analyzed, combined with the low signal to be detected which
is at the limit of the sensitivity of either instruments. In fact the excess we find
above 100 keV is a little bit higher than the $3\sigma$ upper limit given
by Lazzati et al. (2001)\nocite{lazzati01}. Likely, due to its
very low background level and its almost stable behaviour with time
(maximum trend of $\sim 0.5\%$ in the time interval $-113$,$+227$ considered), 
the \sax\ GRBM is more suitable to detect these long duration
signals.
The reliability of the long duration component found by us was also tested
performing the same analysis on $\sim 1000$ background light curves which
did not include any short GRB. No excess was found
in any of the $10$ sets of light curves, each statistically equivalent
to the final sample of light curves which included short GRBs.
If the excess before the trigger time is real, the interpretation
of the excess light curve after the trigger time as only afterglow 
emission is questionable.
A more likely interpretation could be that both the excesses before and 
after the short event are in fact prompt emission released during the``short''
GRBs, which, for its low intensity level, is not singly 
detected. 
This does not exclude that an afterglow component could contribute to
the weak emission we observe. If it was the case, this component would
be very weak and, in any case, much lower than the gamma--ray emission
tail observed by, e.g. Giblin et al. (1999), which was attributed to
afterglow emission.
Data show
a spectral evolution of the long component from soft, before the GRB, to hard 
after it. The significance level of
this hardening is $1.2\cdot10^{-3}$ ($>3\,\sigma$).
As a conclusion, short GRBs could be a special class of long GRBs 
with a strong flash preceded and followed by a very weak emission. 
This makes the short GRBs more intriguing and more fascinating. 
The {\em Swift} mission is expected to provide a robust 
test of the results presented in this paper.

\acknowledgments

We thank Francesco Calura for his invaluable help for testing and
exploiting the response function of the GRBM,
and Davide
Lazzati for useful discussion.
This research was supported by the
Ministry of University and Research of Italy (PRIN 2003020775).

\clearpage

%
%
\begin{deluxetable}{ccccc}
\tablewidth{0pt}
\tablenum{1}
\tablecaption{Mean properties of the long duration component and GRB main event. 
Errors (only due to statistics) are at $1\,\sigma$ confidence level}
\tablehead{
Parameter & GRBs & ${\cal A}=[-19,-2]\,{\rm s}$ & ${\cal B}=[6,16]\,{\rm s}$ & ${\cal C}=[-19,-1]\cup[3,26]\,{\rm s}$
         }
\startdata
$C(40-700)$\tablenotemark{(a)} & $457 \pm 7$ &  $51 \pm 13$ & $32 \pm 10$ & $79 \pm 20$ \\
$C(>100)$\tablenotemark{(a)}  & $374 \pm 7$ &  $27 \pm 13$ & $38 \pm 10$ & $76 \pm 21$ \\
$HR$ & $0.82 \pm 0.01$ & $0.55 \pm 0.13$  & $1.09^{+0.08}_{-0.10}$ & $0.97 \pm 0.08$ \\
$\Gamma$\tablenotemark{b} & $1.30 \pm 0.06$ & $2.5 \pm 0.6$  &  $-1.7^{+1.7}_{-2.9}$ & $0.2^{+0.6}_{-1.0}$ \\
$S(40-700)$\tablenotemark{c} & $6.76 \pm 0.16$ & $0.75^{+0.19}_{-0.18}$  &  $0.46^{+0.21}_{-0.24}$ & $1.30^{+0.33}_{-0.35}$ \\
$F(40-700)$\tablenotemark{d} & $548.\pm 9.$\tablenotemark{e} & $4.4 \pm 1.1$  &  $4.6^{+2.1}_{-2.4}$ & $3.2 \pm 0.8$ \\
\enddata
\tablenotetext{a}{Mean count in counts/burst.}
\tablenotetext{b}{Photon index of the power--law model assumed.}
\tablenotetext{c}{Mean fluence in units of $10^{-7}$~erg/cm$^2$/burst.}
\tablenotetext{d}{Mean flux in units of $10^{-9}$~erg/cm$^2$/s/burst.}
\tablenotetext{e}{Averaged on $1\,$s integration time.}
\label{tab1}
\end{deluxetable}

\clearpage


%
%

\clearpage

%
%

\clearpage

%
%


\begin{thebibliography}{}

\bibitem[Barraud et al. (2003)]{barraud03}
Barraud, C., et al. 2003, \aap, 400, 1021

\bibitem[Belli (1995)]{belli95}
Belli, B.M. 1995, \apss, 231, 43

\bibitem[Boer et al. (2002)]{boer02}
Boer, M., Klotz, A., Atteia, J. L., Pollas, C., and Pinna, H. 2002,
GCN 1408 

\bibitem[Calura et al. (2000)]{calura00}
Calura, F., et al. 2000, AIP Conf. Series, 526, 721 


\bibitem[Connaughton (2002)]{con02} Connaughton, V. 2002, \apj 567, 1028

\bibitem[Costa et al. (1997)]{costa97} Costa, E., Frontera, F., Heise,
  J., et al. 1997, \nat, 387, 783

\bibitem[Costa et al. (1998)]{costa98} Costa, E., et al 1998, 
Adv. Sp. Res., 22, 1129

\bibitem[Feroci et al. (1997)]{feroci97} Feroci, M., et al. 1997 in SPIE Conf. on
EUV, X-Ray, and Gamma-Ray Instrumentation for Astronomy VIII, eds.
O. H. Siegmund \& M. A. Gummin (San Diego: SPIE), Vol. 3114, 186

\bibitem[Frontera et al. (1997)]{frontera97} Frontera, F., et al. 1997, \aaps, 122, 357
%

\bibitem[Ghirlanda et al. 2004]{Ghirlanda04}
Ghirlanda, G., Ghisellini, G. \& Celotti, A. 2004, \aap, 422, 2004

\bibitem[Giblin et al. 1999]{giblin99} Giblin, T. W., et al. 1999,
  \apj, 524, L47

\bibitem[Guidorzi (2002)]{guitesi02} Guidorzi , C., Ph. D. Thesis, 2002, http://www.fe.infn.it/$\sim$guidorzi/

\bibitem[Guidorzi et al. (2004)]{guidorzi04} Guidorzi, C., Montanari,
  E., Frontera, F., et al. 2004, in Proc. 
3rd Rome Workshop on "Gamma Ray Bursts in the Afterglow Era",
ed.s M. Feroci, F. Frontera, N. Masetti, and L. Piro, ASP Conf. Series,
vol. 312, p. 39 
%
%
%
\bibitem[Klotz et al. (2003)]{klotz03} 
Klotz, A., Boer, M., Atteia, J. L. 2003 \aap, 404, 815

\bibitem[Kouveliotou et al. (1993)]{kouveliotou93} Kouveliotou, C., Meegan, C. A., Fishman, G. J.,
et al. 1993, \apj, 413, L101

\bibitem[Lamb et al. 2004]{Lamb04}
Lamb, D.Q. et al. 2004, in Proc. 3rd Rome Workshop on "Gamma 
Ray Bursts in the Afterglow Era", ed.s M. Feroci, F. Frontera, N. Masetti, 
and L. Piro, ASP Conf. Series vol.312, p. 94  

\bibitem[Lazzati et al. (2001)]{lazzati01} Lazzati, D., Ramirez-Ruiz, E., \& Ghisellini, G. 2001, \aap, 379, L39

\bibitem[Montanari et al.(2004)]{montanari04} Montanari, E, Guidorzi,
  C., Frontera, F., et al. 2004, in Proc. 
3rd Rome Workshop on "Gamma Ray Bursts in the Afterglow Era",
ed.s M. Feroci, F. Frontera, N. Masetti, and L. Piro, ASP Conf. Series,
vol. 312, p. 193

\bibitem[Paciesas et al. 2001]{Paciesas01}
Paciesas, W.S. et al. 2004, in in Proc. 2nd Rome Workshop on "Gamma Ray Bursts in 
the Afterglow Era", ed.s E. Costa, F. Frontera, and J. Hijorth (Springer, Berlin)
p. 13

\bibitem[Ricker et al. (2002)]{ricker02}
Ricker, G., et al. 2002, GCN 1399

\bibitem[Smith et al. (1999)]{smith99} Smith, M.J.S., Gandolfi, G.,
  Celidonio, G., et al. 1999, \aaps, 138, 561

\bibitem[Toma et al. (2005)]{toma05}
Toma, K., Yamazaki, R., and Nakamura, T. 2005, \apj, 620, 835

\bibitem[Yamazaki et al. (2004)]{yamazaki04}
Yamazaki, R., Kunihito, I., and Nakamura, T. 2004, \apj, 607, L103

\bibitem[Zhang \& M\'esz\'aros (2004)]{zhang04}
Zhang, B. \& M\'esz\'aros, P. 2004, Journ. Mod. Phys. A, in press
(astro-ph/0311321)


\end{thebibliography}
\end{document}